\documentclass[%
 aip,
 jmp,%
 amsmath,amssymb,
 reprint,%
]{revtex4-2}

\usepackage{graphicx}
\usepackage{dcolumn}
\usepackage{bm}
\usepackage{xcolor}
\usepackage{hyperref}
\usepackage{float}
\hypersetup{
  colorlinks,
  citecolor=blue,
  linkcolor=blue,
  urlcolor=black}

\begin{document}

\preprint{AIP/123-QED}

\title {Superconductivity and Magnetism in Bi-Ni System: From Bulk to Heterostructures}

\author{Gabriel Sant'ana\textsuperscript{1,2}}
\author{Yutao Xing\textsuperscript{3}}
\author{Milton A. Tumelero\textsuperscript{2}}
 \email{matumelero@ig.ufrgs.br}
\affiliation{\textsuperscript{1}Kamerlingh Onnes-Huygens Laboratory, Leiden University, P.O. Box 9504, 2300 RA Leiden, The Netherlands.\\
\textsuperscript{3}Instituto de F\'{i}sica, Universidade Federal do Rio Grande do Sul, 91501-970 Porto Alegre, Brazil.\\
\textsuperscript{2}Instituto de F\'{i}sica, Universidade Federal Fluminense, 24210-346 Niterói, Brazil.}

\date{\today}

\begin{abstract}
Unconventional superconductivity with ingredients of magnetism such as time reversal breaking, triplet pairing and proximity effects has been a long time pursued topic in physics and material science as a new source of innovative phenomena and effects to drive further scientific advances and technologies. The Bi-Ni system has emerged as a convenient candidate for the investigation of such phenomenology. Strictly speaking, the term Bi-Ni system refers to two different schemes, first to the NiBi$_{3}$ intermetallic compound, a 4 K superconductor in which a complex magnetism seems to govern electrical properties in the normal phase, but without implication in the superconducting properties. The second Bi-Ni system is not quite a material but a heterostructure formed by a bilayer of Bi and Ni. Here the superconducting phase has to steal the spotlight by showing strong evidence of time-reversal breaking and triplet pairing featuring a complex unconventional state. The ease of preparing the samples, compared to other candidates for such unconventionality, usually Uranium-based compounds, has accelerated the research in this system. Here, we went through a detailed review of the main results about the electronic properties of these two systems as well as a discussion of the main open questions to be solved in order to reach the complete understanding in this topic.  
\end{abstract}


\maketitle

\section{INTRODUCTION}

The Bismuth–Nickel system has attracted significant attention due to the intriguing interplay of superconductivity, topology, and magnetism that emerges in different material realizations. This rich landscape is exemplified in two distinct contexts that have motivated extensive research. On one hand, the intermetallic compound NiBi$_3$ is a bulk superconductor with evidence in favor of conventional $s$-wave pairing, yet it exhibits puzzling magnetic and transport anomalies whose microscopic origin remains unsettled. On the other hand, Bi/Ni bilayers display superconductivity around 4 K, with experimental reports ranging from conventional Braden-Cooper-Shchiffer (BCS)-like behavior to exotic chiral states, depending strongly on the growth conditions. These contrasting scenarios raise fundamental questions about the role of structure, magnetism, and spin–orbit coupling in enabling superconductivity within the Bi–Ni system.

Here we review the current understanding of both cases. Section~\ref{sec:nibi3_literatura} focuses on NiBi$_3$ single crystals, emphasizing their crystal structure, superconducting and magnetic properties, and unresolved issues in normal-state transport. The subsequent section examines Bi/Ni bilayers, highlighting the role of growth techniques, interfacial phenomena, and spin-related effects in shaping their superconducting state. Together, these two perspectives set the stage for identifying open questions and motivate the investigations carried out in this work.

\section{NiBi$_3$ single crystal: Structure, superconductivity and magnetism properties}
\label{sec:nibi3_literatura}

\begin{figure}[h]
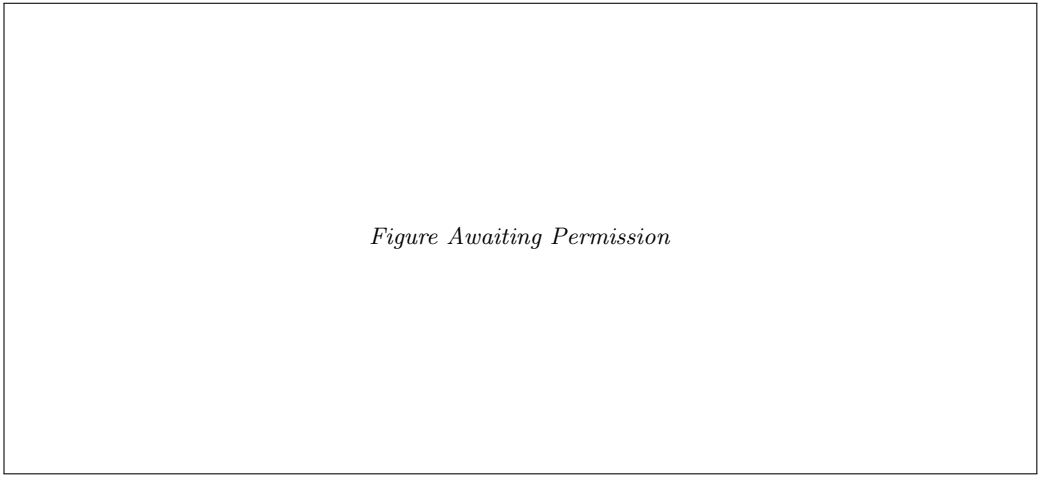

    \centering
    \fbox{
        \parbox[c][6cm][c]{0.95\linewidth}{
            \centering
            \textit{Figure Awaiting Permission}
        }
    }
    \caption{Orthorhombic Structure of NiBi$_3$ along distinct axis. Prepared with Vesta \cite{Vesta}}
    \label{fig:review_nibi3_structure}
\end{figure}

The compound NiBi$_3$ crystallizes in a centrosymmetric orthorhombic structure with space group \textit{Pnma}, consisting of quasi-1D chains of Ni atoms aligned along the \textit{b}-axis. These Ni chains are embedded within a three-dimensional framework formed by Bi atoms, which occupy positions that create distorted octahedral and prismatic coordination around Ni, as schematically shown in Fig. \ref{fig:review_nibi3_structure}. The unit cell contains four formula units, with lattice parameters typically around $a \approx 8.88$~\AA, $b \approx 4.10$~\AA, and $c \approx 11.50$~\AA, though slight variations are reported depending on synthesis conditions \cite{kumar2011physical,silva2013superconductivity}. From a chemical bonding perspective, the crystal structure of NiBi$_3$ can be interpreted as a packing of 1D NiBi$_3$ rods,  primarily governed by strong Ni-Ni and Bi-Ni interactions, whereas Bi-Bi bonding plays only a minor role. The electron localization function further supports this picture, showing delocalized conduction electrons confined within the rods, while valence electrons between the rods remain localized \cite{herrmannsdorfer2011structure}. 

Electron transport in NiBi$_3$ exhibits two main characteristics: an early saturation of the electrical resistivity as a function of temperature at unusually low temperatures, typically in the range of 30–50 K, and a superconducting transition occurring at approximately 4 K.\cite{fujimori2000superconducting,zhu2012surface,nedellec1985anomalous} The latter phenomenon has been the primary focus of research on this material over the past three decades, largely motivated by the possibility of magnetism-related effects.
NiBi$_3$ has been established as a type-II superconductor in the strong-coupling regime\cite{fujimori2000superconducting}, characterized by an electron-phonon coupling constant $\lambda_{e\text{-}p} \approx 0.9$ \cite{zhu2012surface}, an effective mass $m^{*} \approx 10m_{e}$ \cite{fujimori2000superconducting}, and a residual resistivity in the range of 4–8~$\mu\Omega\cdot$cm \cite{fujimori2000superconducting,zhu2012surface,silva2013superconductivity}. The Sommerfeld coefficient is reported to be $\gamma = 12.7$~mJ\,mol$^{-1}$\,K$^{-2}$, while the London penetration depth is approximately 194~nm. \cite{fujimori2000superconducting} The superconducting coherence length is anisotropic, varying from 14 to 38~nm depending on the crystallographic direction \cite{zhu2012surface}, with an average isotropic value of about 26~nm \cite{fujimori2000superconducting}. These parameters result in an upper critical field of approximately 4.7~kOe.\cite{fujimori2000superconducting,zhu2012surface}

\subsection{Magnetic Ordering and Spin-Fluctuations}

Despite containing the ferromagnetic element Ni, the compound NiBi$_3$ exhibits nearly temperature-independent magnetic susceptibility, consistent with the full occupancy of the Ni 3$d$-electron bands and the absence of local magnetic moments \cite{kumar2011physical}. Still, former reports proposed the coexistence of ferromagnetism and superconductivity at temperatures below the superconducting critical temperature (T$_C$) \cite{pineiro2011possible}. Nevertheless, the presence of small ferromagnetic hysteresis within the superconducting state, with the Curie temperature extending far above room temperature, has been attributed to amorphous Ni inclusions inside or at the surfaces of the NiBi$_{3}$ single crystals \cite{silva2013superconductivity}. 
In the Ref. \cite{silva2013superconductivity} at magnetic fields above the upper critical field (B$_{C2}$) exhibit a peculiar behavior characterized by a finite magnetization typical of a ferromagnet. Increasing the temperature above T$_C$ allowed the authors to suppress the superconducting contribution, confirming the ferromagnetic behavior due to nickle inclusions. The saturation magnetization was found to be small and interpreted as arising from residual amorphous Ni inclusions, likely formed during synthesis or due to surface degradation.
Thus, the ferromagnetism is extrinsic to the superconducting state. These Ni-rich amorphous regions are known to exhibit superparamagnetic-like behavior at low temperatures, which can mimic intrinsic ferromagnetic responses.

\begin{figure}
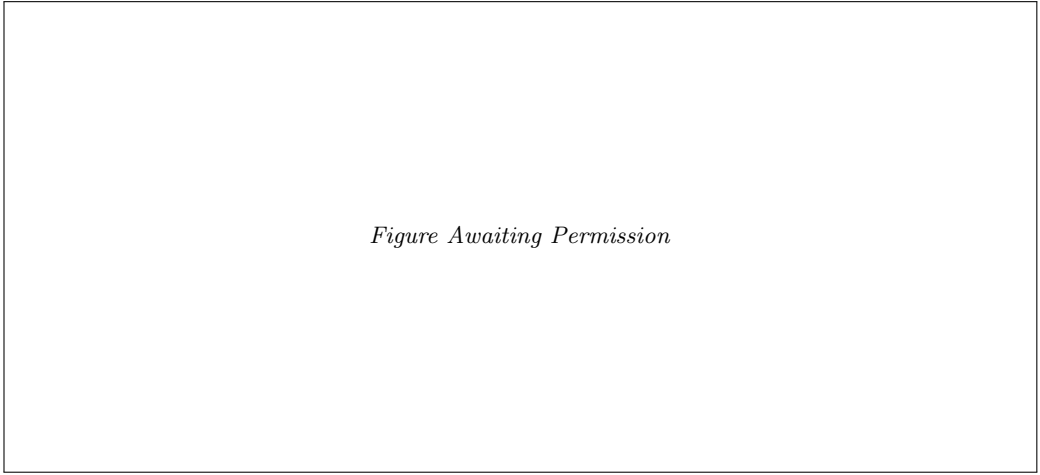

    \centering
    \fbox{
        \parbox[c][6cm][c]{0.95\linewidth}{
            \centering
            \textit{Figure Awaiting Permission}
        }
    }
      \caption{\textbf{Magnetic measurements in NiBi$_3$ single crystals and nanostructured samples.} (a,b) Electron spin resonance (ESR) measurements with the magnetic field applied along the $b$-axis and perpendicular to the $b$-axis, respectively. Adapted from ref. \cite{zhu2012surface} (c) Hall resistivity as a function of magnetic field measured at several temperatures. Adapted from ref. \cite{sant2025linear} (d,e) Magnetic moment as a function of temperature under an applied field of 5 mT and as a function of magnetic field at 1.8~K, respectively. (f) Full magnetic moment versus magnetic field curve for nanostructured samples measured up to 15 T. (g) Magnetic moment as a function of applied magnetic field at different temperatures for a submicrometer structure. Adapted from ref. \cite{herrmannsdorfer2011structure}} 
    \label{fig:nibi3_crystal_magnetism}
\end{figure}

In single crystals samples \cite{huang2024, zhu2012surface,gati2018effect}, the bulk were reported without any long-range magnetic order \cite{zhang2025nodal,silva2013superconductivity}, instead, other types of magnetic signatures were present in the samples such as ferromagnetic fluctuations that were observed by Electron Spin Resonance (ESR) technique \cite{zhu2012surface}. As shown in the dip-like curve manifested up to 160 K for magnetic field applied along the $b$-axis (Fig. \ref{fig:nibi3_crystal_magnetism}\textcolor{blue}{a}) and perpendicular to $b$-axis (Fig. \ref{fig:nibi3_crystal_magnetism}\textcolor{blue}{b}). The vertical black dashed line, at about 0.5 kOe is an eye-guide curve to emphasize the change in the ESR signal from low temperature to 160 K where the ESR peak can be observed. Such a signal is typically associated with correlated spin dynamics and short-range ferromagnetic interactions, rather than isolated paramagnetic behavior. Moreover, the microwave signal of ESR penetrates only a few $\mu$m into the sample, suggesting that ferromagnetic moment fluctuations arise nearly the surface. Furthermore, recent magnetotransport measurements by Sant’Ana \textit{et al.} \cite{sant2025linear} revealed additional evidence of magnetic correlations. As shown in Fig.~\ref{fig:nibi3_crystal_magnetism}{\color{blue}c}, below approximately 150 K
an anomalous Hall contribution emerges. A detailed analysis indicates that this effect is dominated by skew scattering and was attributed to spin fluctuations, consistent with the presence of short-range ferromagnetic correlations in NiBi$_3$.

Additionally, studies have revealed intriguing signatures of spontaneous
magnetism in confinement-sized NiBi$_3$ samples emerging at low temperatures
($T \lesssim 15$ K), which are strongly dependent on synthesis conditions and
microstructural details \cite{herrmannsdorfer2011structure}. Magnetic investigations of submicrometer grains and nanostructured assemblies
show a markedly different and significantly enhanced magnetic response compared
to bulk single crystals, indicating that reduced dimensionality and increased
surface-to-volume ratio play a crucial role in amplifying magnetic signatures (see Fig. \ref{fig:nibi3_crystal_magnetism}{\color{blue}d,e}).
In submicrometer NiBi$_3$ samples, magnetization measurements performed below
$T_C$ reveal a field- and temperature-dependent behavior, as shown in
Figs. \ref{fig:nibi3_crystal_magnetism}{\color{blue}f}: at low fields the system enters the Meissner state, characterized by a reversible diamagnetic response and the absence of hysteresis, whereas at higher fields a clear ferromagnetic hysteresis loop develops and persists into the normal state, signaling spontaneous magnetization coexisting with superconductivity within the same sample volume rather than originating from separate magnetic phases. In nanostructured NiBi$_3$, where the material consists of randomly oriented one-dimensional strands with molecular-scale diameters, magnetic effects are
even more pronounced, as illustrated in Fig. \ref{fig:nibi3_crystal_magnetism}{\color{blue}g}: the saturation magnetization per Ni atom reaches approximately $0.8~\mu_B$ at $\mu_0H$ = 15 T and continues to increase without saturation in pulsed magnetic fields up to 60 T, reaching values as high as $1.5~\mu_B$/Ni. The absence of saturation even at such extreme fields suggests strong magnetic anisotropy and enhanced exchange interactions along the quasi-one-dimensional chains, while the reduced coercivity and remanence observed in the nanostructured samples, compared to the submicrometer case, indicate a softer ferromagnetic character.

These findings challenge the interpretation of ferromagnetism in NiBi$_3$ as purely extrinsic. While amorphous Ni segregation can explain localized magnetic signals in bulk crystals, attempts to magnetic doping with Fe do not indicate any FM ordering \cite{Gonsalves2017FeDoping}. Still, the systematic emergence of ferromagnetic hysteresis in confined geometries points toward a dimensionality-induced magnetic instability, which is supported by the ferromagnetic fluctuations near to the surface. In summary, there has been limited progress in understanding these underlying magnetic phenomena.

\subsection{Superconducting Pairing Symmetry}

\begin{figure}
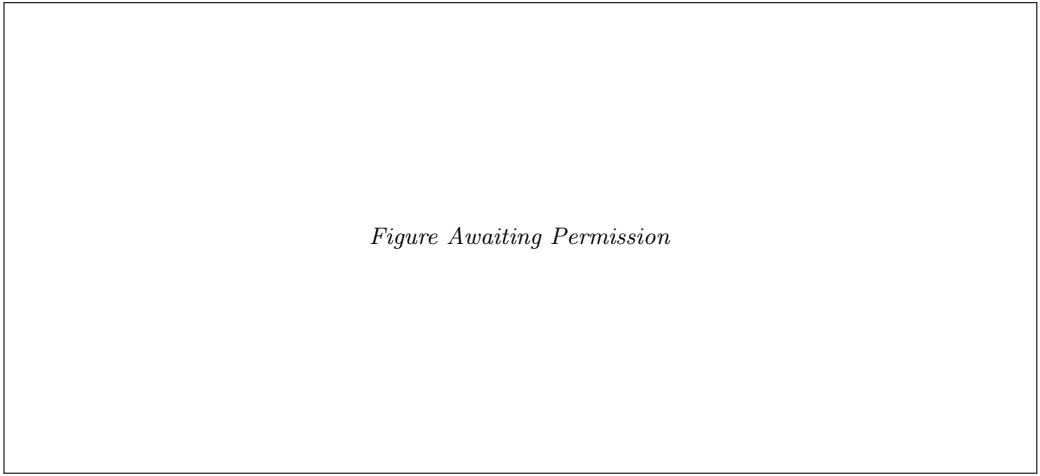

    \centering
    \fbox{
        \parbox[c][6cm][c]{0.95\linewidth}{
            \centering
            \textit{Figure Awaiting Permission}
        }
    }
    \caption{\textbf{Superconducting pairing symmetry in NiBi$_3$.}
    (a–d) Evidence for the absence of time-reversal symmetry breaking in the superconducting state. (a) Polar Kerr effect signal measured across the superconducting transition. Colored symbols correspond to different training-field conditions. Data from ref. \cite{wang2023absence} (b) Zero-field muon spin relaxation (ZF-$\mu$SR) asymmetry as a function of time, showing nearly identical relaxation behavior above and below $T_c$. (c) Temperature dependence of the zero-field Lorentzian relaxation rate $\lambda_{\mathrm{ZF}}(T)$. (d) Temperature dependence of the Gaussian relaxation rate $\sigma_{\mathrm{ZF}}(T)$, associated with static nuclear moments. Adapted from ref. \cite{shang2023fully} (e,f) Conventional superconducting behavior.
    (e) Pressure dependence of the superconducting critical temperature $T_c$. The dashed line represents the BCS-based theoretical description, as indicated in the figure. Adapted from Ref.~\cite{gati2018effect}. (f) Temperature evolution of the superconducting energy gap, following a BCS-like behavior. Adapted from Ref.~\cite{zhao2018singlet}.}
    \label{fig:nibi3_pairing_symmetry}
\end{figure}

Investigations were performed to verify whether the superconducting state of NiBi$_3$ either possesses intrinsic magnetism or coexists with any ferromagnetic phase, by means of advanced high-sensitivity magnetometric techniques such as the magneto-optical Kerr effect (MOKE)~\cite{wang2023absence} and muon-spin rotation and relaxation ($\mu$SR) \cite{shang2023fully}. MOKE measurements were performed using a zero-area-loop Sagnac interferometer in NiBi$_3$ single crystals, which is sensitive to time-reversal symmetry breaking through the detection of out-of-plane magnetization. This optical technique, based on polarized light, can detect quite small spontaneous magnetization through the measurement of polarization changes of few nanoradians. As shown in Fig.~\ref{fig:nibi3_pairing_symmetry}{\color{blue}a}, no increase in the Kerr angle ($\theta_k$) was observed between 2 K and 10 K under both zero-field cooling (ZFC) and training-field protocols. The use of ZFC ensures that the sample enters the superconducting state without any trapped magnetic flux, isolating possible intrinsic magnetic signals from vortex-related artifacts, while the application of a small training field (smaller than $B_{C1}$) tests whether a controlled perturbation can align potential ferromagnetic domains without inducing vortex pinning. The absence of any Kerr signal in both cases indicates that the superconducting transition does not generate a measurable magnetic moment and preserves time-reversal symmetry (TRS).

Similar conclusions have been drawn from $\mu$SR measurements, where the absence of spontaneous magnetic fields below $T_c$ confirms the preservation of time-reversal symmetry. Furthermore, the lack of fast relaxation and/or coherent oscillations in the zero-field $\mu$SR spectra (see Fig. \ref{fig:nibi3_pairing_symmetry}{\color{blue}b}) rules out the presence of magnetic order, either at the surface or in the bulk of NiBi$_3$ single crystals, suggesting that the ferromagnetic signatures reported in earlier studies are most likely of extrinsic origin \cite{shang2023fully}. Additional insight is provided by the temperature dependence of the zero-field relaxation rates shown in Figs. \ref{fig:nibi3_pairing_symmetry}{\color{blue}c,d}. In this framework, the Lorentzian relaxation rate $\lambda_{\mathrm{ZF}}$ is sensitive to slow electronic magnetic fluctuations, while the Gaussian rate $\sigma_{\mathrm{ZF}}$ predominantly reflects static, randomly oriented nuclear moments. The absence of any systematic enhancement in either relaxation channel upon entering the superconducting state indicates that no electronic magnetism or spontaneous internal fields emerge below $T_c$, thereby excluding time-reversal symmetry breaking in NiBi$_3$. In addition, the temperature dependence of the superfluid density in NiBi$_3$ is consistent with a nodeless superconducting gap, well described by an isotropic $s$-wave model.

Further support for conventional superconductivity in NiBi$_3$ is provided by pressure-dependent studies of the superconducting transition temperature $T_c$~\cite{gati2018effect}. As shown in Fig.~\ref{fig:nibi3_pairing_symmetry}\textcolor{blue}{e}, the experimentally observed suppression of $T_c$ under hydrostatic pressure can be quantitatively reproduced within the BCS framework using the relation $T_c = 1.136\Theta_D \exp[-1/VN(E_F)]$. The dashed, dotted, and dash–dotted lines correspond to calculations assuming different pressure dependences of the Debye temperature, namely $d\Theta_D/dp = 0$, $+2.5$, and $+5$ K GPa$^{-1}$, respectively, while the electron–phonon coupling potential $V$ is kept pressure independent. In all cases, the pressure-induced reduction of the electronic density of states at the Fermi level, $N(E_F)$—experimentally determined to decrease approximately linearly with pressure—accounts for the monotonic decrease of $T_c$ in good agreement with both resistivity and magnetization data. The comparable agreement obtained for all assumed $\Theta_D(p)$ scenarios indicates that changes in $N(E_F)$, rather than phononic stiffening, provide the dominant contribution to the suppression of superconductivity under pressure, consistent with a weak-coupling BCS description.

Complementary evidence for a conventional pairing state in NiBi$_3$ is obtained from direct spectroscopic probes of the superconducting gap. Andreev point-contact spectroscopy performed along several crystallographic directions reveals a fully developed, nodeless gap whose temperature dependence closely follows the BCS prediction, as shown in Fig.~\ref{fig:nibi3_pairing_symmetry}\textcolor{blue}{f}. Measurements using a normal-metal Au tip exhibit a pronounced proximity effect, whereas the use of a spin-polarized La$_{0.7}$Sr$_{0.3}$MnO$_3$ (LSMO) tip leads to a strong suppression of Andreev reflection due to the exchange field, in agreement with a spin-singlet pairing state in NiBi$_3$~\cite{zhao2018singlet}. These findings are further supported by specific-heat measurements, which display a superconducting jump and low-temperature behavior consistent with BCS expectations for an isotropic $s$-wave gap \cite{fujimori2000superconducting,zhu2012surface,wang2023absence}. Taken together, spectroscopic and thermodynamic results consistently point toward phonon-mediated, spin-singlet superconductivity in NiBi$_3$. 

Despite this strong results point to a conventional superconducting s-wave ground state of NiBi$_3$, quite BCS-like, and the clear manifestation of magnetic effects - either from bulk, surface effects or size confinement - the relation between this two phenomena remains unclear. The magnetic fluctuations picture could be good description to the magnetic effect above the critical temperature. Near the T$_C$, no direct evidences of fluctuation was observed by muon spectroscopy \cite{shang2023fully}, both in normal and SC phase. However, at the SC side of the transition, an abnormal pinning dependence of magnetic field and a reentrance in the vortex liquid phase at low field, spanning an high Ginzburg-Levanyuk parameter indicate a strong dependence from disorder, which could be related to magnetic disorder \cite{Rollano2023}. Undoubtedly, this system deserves further attention.

\subsection{Resistivity and Magnetotransport Anomalies}

\begin{figure}
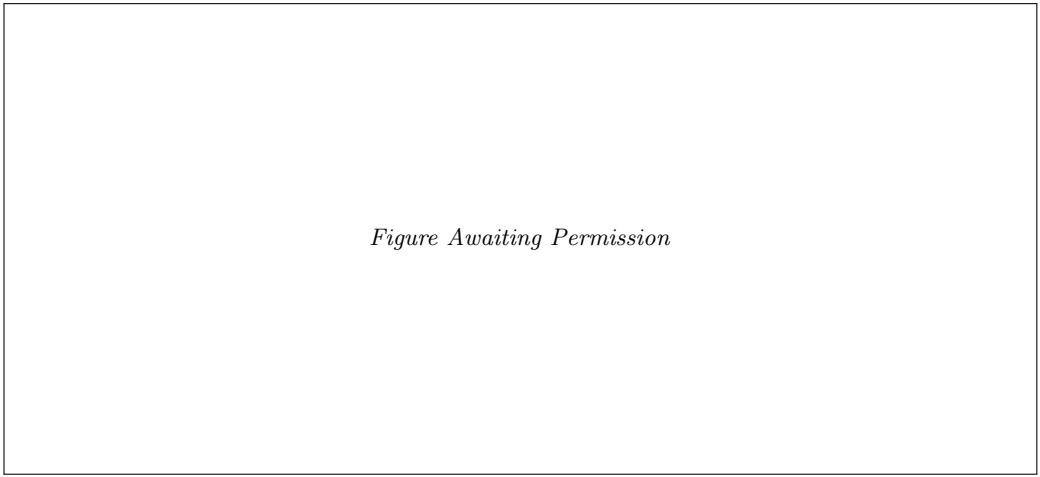

    \centering
    \fbox{
        \parbox[c][6cm][c]{0.95\linewidth}{
            \centering
            \textit{Figure Awaiting Permission}
        }
    }
    \caption{\textbf{Resistivity and Magnetotransport Anomalies.} (a) Longitudinal resistivity as a function of temperature. \textit{Inset}: zoomed-in view highlighting the superconducting transition. Adapted from Ref. \cite{siva2015spontaneous}. (b) Photograph of a NiBi$_3$ single crystal. (c) First derivative of the longitudinal resistivity. The colored regions indicate the superconducting (SC) phase, the normal state described by the Bloch-Gr\"uneisen (BG) model, and the deviation from BG behavior. The solid line corresponds to a simulation based on the BG equation. (d) Longitudinal magnetoresistance as a function of magnetic field at several temperatures. Adapted from Ref.~\cite{sant2025linear}. (e,f) Kohler's rule analysis of the magnetoresistance at different temperatures. Adapted from ref. \cite{zhu2012surface}. (g,h) Normalized quadratic and linear coefficients as functions of temperature, extracted from polynomial fits to the magnetoresistance in the high- and low-field regimes, respectively.}
    \label{fig:nibi3_anomalies_resistivity}
\end{figure}

Another interesting feature NiBi$_{3}$ that has been addressed in the past is the temperature dependence of the longitudinal resistivity $\rho_{xx}(T)$ (Fig. \ref{fig:nibi3_anomalies_resistivity}{\color{blue}a}). Usually, resistivity in metals follows the the conventional Bloch–Grüneisen (BG) description 
\begin{equation}
    \rho_{BG}(T) = \rho_{0} + A\left(\frac{T}{\Theta_{D}}\right)^{5}\int_{0}^{\frac{\Theta_{D}}{T}}\frac{x^{5}\mathrm{d}x}{(e^{x} - 1)(1 - e^{-x})},
\end{equation}
where $\Theta_D$ is the Debye temperature and $A$ is a material-dependent constant, which describes electron–phonon scattering. According to the BG model, the first derivative of resistivity initially increases as $T^4$, reaching a maximum that marks the crossover where phonon population becomes significant. Beyond this point, the derivative slightly decreases and approaches a plateau, marking the regime where electron-phonon scattering saturates. In NiBi$_3$, on the other hand, an anomaly appear as a peak in the resistivity first derivative at low temperature and deviates from BG predictions (solid line in Fig. \ref{fig:nibi3_anomalies_resistivity}{\color{blue}c}). Additionally, an early saturation of resistivity, where $\rho_{xx}$ reaches a plateau at high temperatures, well before the phonon-scattering limit predicted by the BG model is also observed. Various studies have proposed alternative mathematical models to fit the $\rho_{xx}(T)$ curve in order to capture its anomalous behavior, ranging from parallel resistor models \cite{fujimori2000superconducting, nedellec1985anomalous} to exponential contributions \cite{zhu2012surface}, as well as modifications of the BG prediction by including additional scattering mechanisms \cite{shang2023fully}. Such an effect has been extensively studied in certain intermetallic compounds, notably those with the A-15 crystal structure
\footnote{The term A15 refers to the $\beta$-tungsten-type cubic structure, with formula A$_3$B, where atoms A form linear chains along the cubic axes. This geometry often leads to strong electron–phonon coupling and high $T_C$ superconductivity \cite{dew1975superconducting}}. 
In such materials, early saturation is commonly attributed to the Ioffe–Regel limit, where the electronic mean free path $\ell_{mfp}$ approaches the interatomic spacing $a$, effectively reaching the shortest possible scattering length for charge carriers. Beyond this limit, further temperature increase no longer reduces $\ell_{mfp}$, leading to a resistivity that no longer grows with $T$. Particularly, for NiBi$_3$ the resistivity become saturated in between 150-200 K, where the change in curvature become pronounced. Still, a complete understanding of the underlying scattering mechanisms remains elusive. 

Interestingly, in the same temperature range where the Bloch–Grüneisen description breaks down, the longitudinal magnetoresistance (MR) develops unconventional features that deviate from ordinary metallic behavior. As shown in Fig.~\ref{fig:nibi3_anomalies_resistivity}\textcolor{blue}{d}, at low temperatures the MR is purely parabolic in magnetic field, consistent with conventional orbital magnetoresistance. Upon increasing temperature beyond the peak observed in $\mathrm{d}\rho/\mathrm{d}T$, the MR gradually evolves into a sigmoidal field dependence, characterized by a pronounced linear contribution at low fields (up to $\sim$1 T), followed by a quadratic regime at higher fields. Polynomial analysis of the MR curves, summarized in Fig.~\ref{fig:nibi3_anomalies_resistivity}\textcolor{blue}{g,e}, reveals that the normalized quadratic coefficient decreases with temperature, while the linear coefficient grows and scales with the longitudinal resistivity~\cite{sant2025linear}.

This coexistence of linear and quadratic MR components is further reflected in a clear violation of Kohler’s rule, whose validity relies on the assumption of a single conduction channel with isotropic, field-independent scattering rates \cite{kohler1938magnetischen}. Indeed, Kohler plots constructed from the experimental data show that scaling holds only at low temperatures, whereas significant deviations emerge above $\sim$60 K \cite{zhu2012surface}, indicating the onset of additional scattering mechanisms. In this regime, transport can no longer be described by a single relaxation time, suggesting that distinct scattering processes govern zero-field resistivity and magnetotransport. Recently, spin fluctuations have been proposed as a possible origin of this additional scattering channel, providing a unified framework to connect the anomalous temperature dependence of $\rho_{xx}(T)$ with the emergence of linear magnetoresistance \cite{sant2025linear}. Nevertheless, the microscopic origin of these anomalies and their relation to the electronic structure of NiBi$_3$ remain open questions.

\subsection{Topology in the Band Structure}

\begin{figure}[h]
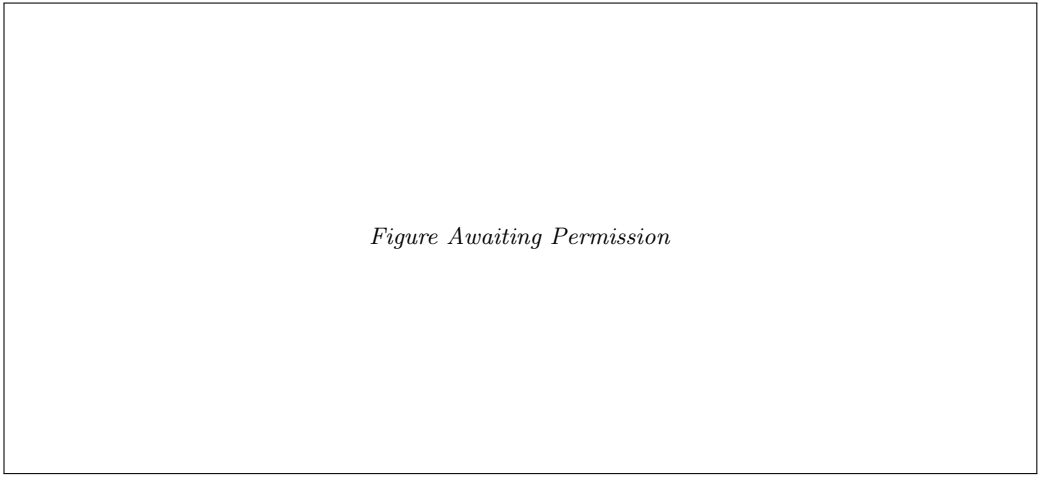

    \centering
    \fbox{
        \parbox[c][6cm][c]{0.95\linewidth}{
            \centering
            \textit{Figure Awaiting Permission}
        }
    }
    \caption{\textbf{ARPES Dispersion Maps.} (a,b) Measurements along the $\Gamma \rightarrow X$ direction at photon energies of 6.7 eV and 6.2 eV, respectively.  (c) Measurement along the $\Gamma \rightarrow Y$ direction at the same photon energies. (d) Momentum–energy points obtained by fitting Gaussian peaks to the momentum-dispersive curves in panels (a) and (b). (e,f) DFT band structure calculations along $\Gamma \rightarrow X$ and $\Gamma \rightarrow Y$, respectively. Data from ref. \cite{adriano2023bulk}}
    \label{fig:review_nibi3_tss}
\end{figure}

The electronic structure of NiBi$_3$ exhibits a rich and spin-orbit coupling (SOC)-dependent topology. In the absence of spin–orbit coupling (SOC), the bulk band structure hosts multiple line degeneracies, including nodal loops and nodal chains. When SOC is introduced, the nodal loops enclosing the valence bands become fully gapped, resulting in a bulk gap that supports $Z_2$ topological surface states. For other band pairs, SOC gives rise to a three-dimensional Dirac point formed by the top valence and bottom conduction bands \cite{zhang2025nodal}. These topological features lie slightly below the Fermi energy, suggesting that modest tuning (e.g., by doping or applying pressure) could shift them to $E_F$ and make them relevant for transport properties. Together, these results indicate a complex interplay between line-node physics and $Z_2$ topology in NiBi$_3$. Experimental evidence supports these theoretical predictions. Angle-resolved photoemission spectroscopy (ARPES) measurements reveal linear energy–momentum dispersions consistent with topological surface states along the $\Gamma \rightarrow X$ (Fig. \ref{fig:review_nibi3_tss}\textcolor{blue}{a,b}) and $\Gamma \rightarrow Y$ (Fig. \ref{fig:review_nibi3_tss}\textcolor{blue}{c}) directions \cite{adriano2023bulk}. Along $\Gamma \rightarrow X$, a Dirac-like cone manifests a linearly dispersing feature, while along $\Gamma \rightarrow Y$ a saddle-like topological surface state is observed. Figure~\ref{fig:review_nibi3_tss}\textcolor{blue}{d} shows the dispersion points obtained by fitting momentum–distribution curves for different energies using a Gaussian function in the TSS region of the $\Gamma \rightarrow X$ maps. For comparison, Figs.~\ref{fig:review_nibi3_tss}\textcolor{blue}{e} and \ref{fig:review_nibi3_tss}\textcolor{blue}{f} present the corresponding DFT band structure calculations along the same directions.

In conclusion, it seems that the NiBi$_{3}$ host, in the bulk, a BCS-like superconductor state, however, the quite robust evidences of magnetic fluctuations and topology in the normal-state properties bring some very exotic physics to be discussed in this system. Some interesting features that are not clear and deserves future research is the relation of the surface topology and the magnetic fluctuation. And further, if the superconductivity at the surface of the NiBi$_{3}$ spam the same symmetries as the bulk or if it is modified by the topological surfaces states and magnetic fluctuations.

\subsection{Thin Films and Bi-Ni interdiffusion}

NiBi$_3$ has been synthesized in thin-film form, usually by co-deposition techniques \cite{park2024superconducting,Das2023} or by stacking thin layers of Bi and Ni \cite{bhatia2018superconductivity}. In multilayers, the formation of NiBi$_3$ occurs due to a reaction–diffusion mechanism near the interface \cite{liu2021spontaneous,sant2024ni}. Interestingly, the $T_C$ in thin films has been related to small deviations in the stoichiometry of the NiBi$_3$ compound, reaching a maximum value at approximately Ni$_{0.21}$Bi$_{0.79}$ \cite{park2024superconducting}. This behavior was correlated with the carrier density and the residual resistivity ratio which are also maximum at this composition. In thin films, the presence of segregated and unreacted Ni is rather probable; many reports present magnetic properties in these samples, with hysteresis curves at low temperatures that extend above room temperature \cite{Das2022-xas,bhatia2018superconductivity}.
.

Another Bi–Ni binary system usually found in thin films is the NiBi compound, also reported as a superconductor. NiBi usually appears near the interface between NiBi$_{3}$ and Ni in thin films \cite{liu2018superconductivity, sheikhi2018growth}, as well as in nanoparticles \cite{smuda2020}. Although no experimental report of magnetism in this system has been found, to the best of our knowledge, this phase presents a superconducting transition near 4.25 K, which is higher than that of NiBi$_3$ \cite{fujimori2000superconducting}. The critical field of NiBi is also higher than the NiBi$_3$ \cite{LIU2020}. However, only a few studies discuss this topic. Liu et al. \cite{liu2018superconductivity} curiously found the formation of a NiBi layer at the Bi/NiBi$_3$ interface rather than at the Ni/NiBi$_3$ interface. Park et al. \cite{park2024superconducting} prepared the NiBi phase using a co-evaporation method; however, no superconducting transition was observed in this sample, which could be related to the crystal structure. Theoretical studies seem to indicate that ferromagnetic interactions could coexist with superconductivity \cite{SARLI2019}.

\section{Bi/Ni bilayer: Growth Techniques, Superconducting Properties and Spin-Related Phenomena}

Individually, neither Ni nor Bi exhibits superconductivity under ambient conditions. Nickel is a ferromagnet and therefore does not support a superconducting condensate, while bismuth becomes superconducting only under high pressure~\cite{li2017pressure}, in its amorphous form, or at ultra-low temperatures~\cite{shier1966}. Remarkably, when combined in a bilayer geometry, a superconducting transition emerges with a critical temperature $T_C \sim 4$ K. This unexpected behavior was first reported by Moodera \textit{et al.}~\cite{moodera1990superconducting} and later systematically studied by LeClair \textit{et al.}\cite{leclair2005coexistence}, who fabricated Bi/Ni bilayers by ultra-high-vacuum evaporation at low deposition rates onto cryogenically cooled substrates. For structures up to Bi(40 nm)/Ni(4 nm), these studies revealed that Bi grows in a metastable face-centered-cubic (FCC) phase on Ni/Al$_2$O$_3$, in contrast to its bulk rhombohedral structure. The presence of the Ni layer is crucial for stabilizing this metastable phase, which has been proposed as a key ingredient for the emergence of superconductivity in the bilayer.

Tunneling spectroscopy on Al/Al$_2$O$3$/Ni/Bi junctions further demonstrated that superconducting correlations penetrate into the ferromagnetic Ni layer, with an estimated coherence length $\xi{_\mathrm{Ni}} \sim 3$–4 nm, indicating a proximity effect that extends beyond a simple interfacial phenomenon. Moreover, measurements performed from both sides of the bilayer confirmed that superconductivity is not confined to the Bi/Ni interface but persists throughout the Bi layer for thicknesses up to at least 40 nm. Taken together, these observations already suggest that superconductivity in Bi/Ni bilayers cannot be fully understood within the framework of conventional superconductor/ferromagnet proximity effects, motivating subsequent investigations into the possible emergence of unconventional and symmetry-broken superconducting states. 

\subsection{Evidences of Chiral Superconductivity}

\begin{figure}[h]
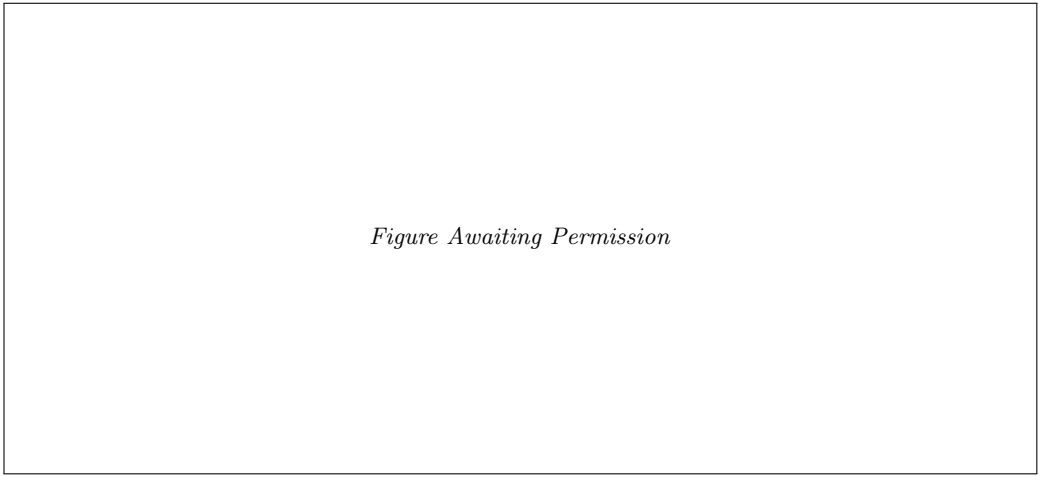

    \centering
    \fbox{
        \parbox[c][6cm][c]{0.95\linewidth}{
            \centering
            \textit{Figure Awaiting Permission}
        }
    }
    \caption{\textbf{Experimental Signatures of Unconventional Superconductivity in Bi/Ni Bilayers.} (a) RHEED and TEM images, highlighting the epitaxial growth and sharp interfaces of Bi and Ni. Cooper pairs are schematically shown at the Bi surface. (b) Critical temperature (T$_C$) as a function of Bi thickness for several Bi/Ni bilayers. (c,d) Conductance curves as a function of applied voltage obtained by Point-Contact Andreev Spectroscopy for a Bi(20 nm)/Ni(2 nm) sample using (c) Au and (d) LSMO tips. The inset in (c) shows the expected conductance for a conventional superconductor (Nb) with an Au tip. (e,f) Kerr signal as a function of temperature for a Bi(25 nm)/Ni(2 nm) sample measured along the (e) Bi surface and (f) Ni/MgO interface.  Data in (a)-(d) are from \cite{gong2017time}, while data in (e) and (f) are from \cite{gong2015possible}.}
    \label{fig:review_bini_kerr_andreev}
\end{figure}

Following these observations of an exotic superconducting state along the entire bilayer, several studies investigated Bi/Ni bilayer epitaxial growth by molecular beam epitaxy (MBE) using cryogenic temperatures for Bi deposition (110 K)\cite{gong2015possible,wang2017anomalous,gong2017time,chauhan2019nodeless}. Structural characterization such as Reflection High Energy Electron Diffraction (RHEED) and Transmission Electron Microscopy (TEM) images provides compelling evidence of epitaxial growth and sharp interfaces, ruling out the formation of Bi-Ni alloy at the interface, as shown in Fig. \ref{fig:review_bini_kerr_andreev}\textcolor{blue}{a}. In these samples, Gong et al. concluded that the interface triggers superconductivity but is not the location of the condensate itself. Their reasoning is based on the observation that superconductivity reappears in Bi layers far from the interface, even though the interface is unchanged (see saturation of T$_C$ in Fig. \ref{fig:review_bini_kerr_andreev}\textcolor{blue}{b}). This indicates that the interface modifies conditions in Bi to enable superconductivity, but the Cooper pairs actually form throughout the bilayer rather than being confined to the interface, as previous argued by LeClair and Moodera. Fundamentally, superconductivity is different from conventional S/F proximity effects. 

Moreover, unconventional superconducting features were also observed in these samples. Remarkable insights emerged from point-contact Andreev-reflection spectroscopy on Bi(25 nm)/Ni(2 nm) specimens. For a conventional $s$-wave superconductor probed with a normal-metal tip (e.g., Au), the differential conductance typically shows coherence peaks at $\pm\Delta$ and a suppressed subgap conductance at zero bias in the tunneling (large-barrier) limit, as in the 100 nm Nb film shown in the inset of Fig. \ref{fig:review_bini_kerr_andreev}\textcolor{blue}{c}. In the Bi/Ni bilayer, however, a single peak centered at zero bias is observed from 1.43 K to 4.71 K when measured with an Au tip, exceeding the behavior expected for conventional $s$-wave pairing, as presented in Fig. \ref{fig:review_bini_kerr_andreev}\textcolor{blue}{c}. Even more strikingly, using a spin-polarized LSMO tip, the zero-bias peak remains essentially unsuppressed. For spin-singlet Cooper pairs, exchange splitting at a ferromagnetic interface reduces strongly the Andreev reflection, leading to diminished subgap conductance \cite{gong2015possible}. Theoretical discussion suggested the presence of Anderson-Brinkman-Morel state in these bilayer \cite{he2022evidence,shang2020}. Overall, the robustness of the zero-bias peak against normal and strong spin polarization tip points to unconventional superconductivity, in particular, compatible with equal-spin triplet, such as $p$-wave, pairing in Bi/Ni. 

Supporting this conclusion, Polar Kerr Effect measurements with a Sagnac interferometer reported a spontaneous breaking of TRS at the superconducting transition of the Bi(25 nm)/Ni(2 nm) sample (see pink symbols in Fig. \ref{fig:review_bini_kerr_andreev}\textcolor{blue}{e}). The technique is highly sensitive to out-of-plane magnetic moments at the surface and the increase in $\theta_K$ is observed in the same temperature range of the superconducting transition (dark cyan symbols in Fig. \ref{fig:review_bini_kerr_andreev}\textcolor{blue}{e}). This increase is only detected on the Bi surface, whereas the Ni/MgO side shows no Kerr signal.

Essentially, the spontaneous Kerr rotation onsets at $T_C$, strongly suggesting that TRS breaking is generated by superconductivity in epitaxial Bi/Ni bilayers. In the original work \cite{gong2017time}, Gong \emph{et al.}  proposed a topological scenario based on a $d\!+\!id$ order parameter mediated by ferromagnetic fluctuations in Ni. However, later experiments with time-domain terahertz spectroscopy technique by Chauhan \emph{et al.} \cite{chauhan2019nodeless} in epitaxial Bi(10 nm)/Ni(1 nm) sample shows a nodeless, fully gapped superconducting state. The temperature dependence of the gap and the superfluid spectral weight follow conventional BCS-like behavior, effectively ruling out any $d$-wave or odd-frequency pairing, and indicating that superconductivity extends across the entire bilayer rather than residing solely in Bi. Combined with strong spin–orbit coupling and TRSB, these observations are consistent with a nodeless chiral $p$-wave–like state ($p_x+ip_y$).

Parallel to these findings, studies by Wang \emph{et al.} \cite{wang2017anomalous} also demonstrated evidence for chiral $p$-wave superconducting states in epitaxial Bi(30 nm)/Ni(2 nm) bilayers. Using a specially designed hybrid SQUID (Bi/Ni-Pb), the authors detected anomalous magnetic moments through hysteretic interference patterns in the superconducting state. These observations, which remained robust against in-plane magnetic fields (ruling out ferromagnetic origins), were attributed to chiral domain dynamics and incomplete cancellation of edge currents. Magnetization measurements in these bilayers shows a non-usual behavior, with the presence of stray field due to the nickel layer \cite{ZHOU2017}.

\subsection{Formation of a NiBi$_3$ interlayer}

\begin{figure}
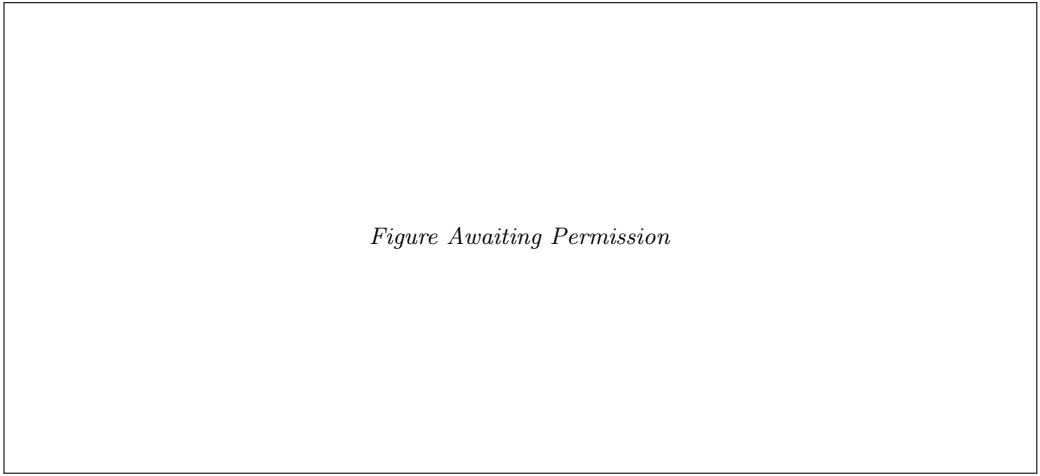

    \centering
    \fbox{
        \parbox[c][6cm][c]{0.95\linewidth}{
            \centering
            \textit{Figure Awaiting Permission}
        }
    }
    \caption{\textbf{Spontaneous Formation of the NiBi$_3$ compound.} (a) High-Resolution Transmission Electron Microscopy images for (a) Bi(38 nm)/Ni(20 nm). Adapted from ref. \cite{liu2018superconductivity} (c) Magnetic moment normalized by area as a function of annealing time. Measurement were done with Polarized Neutron Reflectometry (PNR) and DC SQUID. \textit{Inset:} normalized magnetic moment as a function of magnetic field. (c) Normalized peak intensity from the Grazing-incidence XRD measurements plotted as a function of annealing time at 70 $^\circ$C. Data from ref\cite{vaughan2020origin}. (d) Normalized resistance as a function of temperature for Ni(6 nm)/Bi(40 nm) bilayers, measured immediately after growth and after 24 days of storage at room temperature. Data taken from ref. \cite{hayashi2024two}}. 
    \label{fig:bini_formation_nibi3}
\end{figure}

On the other hand, despite strong evidence for unconventional superconductivity compatible with chiral $p$-wave symmetry extending over the entire bilayer, several studies have reported an alternative explanation based on the spontaneous formation of Bi–Ni intermetallic compounds. In particular, for non-epitaxial grown made by sputtering or pulsed laser deposition (PLD), characterization data unambiguously demonstrate the spontaneous formation of NiBi$_3$ at the Bi/Ni interface during deposition \cite{sant2024ni,vaughan2020origin,liu2018interface,liu2021spontaneous,hayashi2024two}. The superconducting transition temperature observed in these bilayers is remarkably close to that of bulk NiBi$_3$ ($T_C \sim 4$ K). As shown in Fig. \ref{fig:bini_formation_nibi3}{\color{blue}a}, high-resolution TEM images for Bi(38 nm)/Ni(20 nm)/SiO$_2$ clearly reveal the presence of the NiBi$_3$ compound irrespective of Ni thickness.

An intriguing aspect of these non-epitaxial Bi/Ni bilayers is the nontrivial pathway leading to the formation of the NiBi$_3$ interlayer. Systematic structural and magnetic studies reveal that the reaction kinetics depend sensitively on both the Bi and Ni thicknesses. For relatively thick Ni layers, the available Bi is fully consumed, resulting in the formation of mixed NiBi and NiBi$_3$ phases at the interface \cite{liu2018interface}. Conversely, for a fixed Ni thickness, increasing the Bi layer thickness promotes the complete consumption of Ni and leads to the growth of a thicker and more homogeneous NiBi$_3$ layer \cite{vaughan2020origin,sant2024ni}.

The kinetics of this reactive process have been investigated in detail using time-resolved magnetic and structural probes \cite{vaughan2020origin}. As shown in Fig.~\ref{fig:bini_formation_nibi3}\textcolor{blue}{b}, magnetization measurements performed by SQUID magnetometry and polarized neutron reflectometry (PNR) reveal a gradual reduction of the Ni magnetic moment upon annealing, consistent with the progressive consumption of ferromagnetic Ni. Complementary grazing-incidence X-ray diffraction measurements, displayed in Fig.~\ref{fig:bini_formation_nibi3}\textcolor{blue}{c}, show a concomitant decrease in the Bi Bragg peak intensity together with the emergence and growth of NiBi$_3$ diffraction peaks, directly tracking the conversion of elemental Bi into the intermetallic compound. The normalized peak intensities demonstrate that the growth of the NiBi$_3$ layer proceeds in tandem with the depletion of both parent elements. These observations indicate that the formation of NiBi$_3$ is not governed by a simple diffusion-limited process, but instead reflects a reaction kinetics controlled by the mutual availability of Bi and Ni at the interface.

A more comprehensive picture of the nucleation and growth of NiBi$_3$ in Bi/Ni bilayers was later provided by Liu et al. \cite{liu2021spontaneous}, who highlighted the critical role of deposition sequence. When Bi was deposited first, a continuous and relatively smooth NiBi$_3$ layer formed at the interface. In contrast, when Ni was deposited first, NiBi$_3$ nucleated as three-dimensional nanoparticles, often with vertical dimensions exceeding the nominal Bi and Ni thicknesses. These contrasting morphologies have important consequences for strain, defect density, and superconducting properties. Nevertheless, the $T_C$ barely changed. In both cases, the presence of NiBi$_3$ appears to be the key factor for the manifestation of superconductivity. The role of NiBi$_3$ was investigated in the Bi/Ni/Bi trilayer structured prepared by thermal evaporation at 4.2 K. Interestingly, no signal of superconductivity was observed, mainly due to the suppression of interdiffusion and thus no NiBi$_3$ was formed \cite{liu2018interface}. Another striking example is presented in Fig.\ref{fig:bini_formation_nibi3}\textcolor{blue}{d}, where a Ni(6 nm)/Bi(40 nm)/SiO$_2$ bilayer showed no superconducting transition immediately after growth. However, after 24 days of storage at room temperature and ambient pressure, a sharp transition with $T_C \sim 4$ K emerged, consistent with the gradual formation of NiBi$_3$ at the interface \cite{hayashi2024two}. 
 
This behavior contrasts sharply with epitaxial Bi/Ni bilayers grown by molecular beam epitaxy, where Ni is deposited at room temperature and Bi at cryogenic temperatures ($\sim$110 K)~\cite{gong2015possible,gong2017time,wang2017anomalous,chao2019superconductivity,cai2023nonreciprocal}. In those systems, no intermetallic compound is detected, despite the observation of a comparable superconducting transition temperature. Nevertheless, a recent study by Sant’Ana \textit{et al.}~\cite{sant2025violation} highlighted the critical role of the Bi growth temperature. In that work, Bi(10 nm)/Ni(1 nm) bilayers were grown by MBE on MgO and Al$_2$O$_3$ substrates with Bi deposited at room temperature, resulting in a fully reactive interface and the formation of an interface-free NiBi$_3$ layer with a thickness of approximately 6 nm. The reported $T_C$ was about $\approx 3.3$ K, slightly reduced compared to bulk NiBi$_3$, consistent with thickness-induced suppression in the quasi-two-dimensional limit.

Regarding the superconducting properties of these non-epitaxial Bi/Ni bilayers, the thickness dependence of $T_C$ for a fixed Ni layer and varying Bi thickness follows the same trend as epitaxial films (see Fig. \ref{fig:review_bini_kerr_andreev}\textcolor{blue}{c}), with saturation at higher Bi thicknesses reflecting the approach to bulk NiBi$_3$ \cite{liu2018interface,vaughan2020origin,Doria_2022}. Small T$_C$ oscillations by changing the NiBi$_3$ thickness has also been reported \cite{Doria_2022}. Consistently, the magnetic response for in-plane and out-of-plane fields reproduces the expected bulk-like behavior with anisotropy due to the geometrical aspects. Naturally, the corresponding coherence length ($\xi$) also approaches that of bulk NiBi$_3$.

\subsection{Unconventional Superconductivity Features in NiBi$_3$}

Unconventional transport phenomena have been recently reported in NiBi$_3$ originated from Bi/Ni systems. Among them, nonreciprocal superconducting transport, which manifested as a finite second-harmonic resistance component that reverses sign under magnetic-field inversion. Such effects have been observed both in epitaxial Bi/Ni bilayers \cite{cai2023nonreciprocal} and in non-epitaxial Bi/Ni heterostructures involving interfacial NiBi$_3$~\cite{hayashi2024two}. As illustrated in Fig. \ref{fig:unconventional_nibi3}\textcolor{blue}{a}, in the normal state at $T=5$ K both the first-harmonic longitudinal resistance $R_x$ and the anti-symmetrized second-harmonic signal $R_{2x}$ are essentially field independent, with $R_{2x}$ remaining below the experimental noise level. Upon entering the superconducting transition region ($T \approx 4$ K), $R_x$ develops a large symmetric magnetoresistance due to field-induced suppression of superconductivity, while a pronounced antisymmetric $R_{2x}$ signal emerges. The second-harmonic response exhibits extrema at finite magnetic fields and reverses sign upon field inversion, demonstrating its nonreciprocal character and its intimate connection to the superconducting state.

Nonreciprocal transport requires the simultaneous breaking of inversion and time-reversal symmetries and is widely regarded as a hallmark of strong SOC and enables current-direction-dependent responses. In the Bi/Ni bilayer geometry, the absence of any detectable first- or second-harmonic field dependence in the normal state, together with the sharp onset of a finite $R_{2x}$ signal in the superconducting transition, indicates that the nonreciprocity is not a trivial normal-state magnetotransport effect but instead emerges from the superconducting condensate itself. This behavior is naturally interpreted in terms of strong interfacial SOC of Rashba-like origin, arising from the structural inversion asymmetry intrinsic to Bi/Ni heterostructures. Moreover, the strong enhancement of the nonreciprocal signal near $T_c$ suggests that the relevant energy scale governing transport is reduced from the Fermi energy to the superconducting gap, thereby amplifying the relative influence of SOC and magnetic field on the electronic response.

\begin{figure}
    \centering
    \fbox{
        \parbox[c][6cm][c]{0.95\linewidth}{
            \centering
            \textit{Figure Awaiting Permission}
        }
    }
    \caption{\textbf{Unconventional superconducting properties in NiBi$_3$ thin films.} (a) Nonreciprocal superconducting transport. The first (upper panel) and second harmonic (lower panel) magnetoresistance, R$_x$ and R$_{2x}$, as a function of the in-plane magnetic field B (perpendicular to the current) at T = 4 K and 5 K for Bi(40)/Ni(6)/SiO$_2$, adapted from \cite{hayashi2024two} (b-d) Magnetotransport measurements in parallel and perpendicular field configurations for Bi(10 nm)/Ni(1 nm)/Al$_2$O$_3$. (e) In-plane upper critical field divided by the Clogston-Chandrasekar limit (B$_P$) as function of reduced temperature for Bi(10 nm)/Ni(1 nm)/Al$_{2}$O$_{3}$ (S\#1 and S\#2) and Bi(10 nm)/Ni(1 nm)/MgO. Adapted from \cite{sant2025violation}}
    \label{fig:unconventional_nibi3}
\end{figure}

Nevertheless, an important and complementary perspective has recently emerged from the work of Sant’Ana \textit{et al.}~\cite{sant2025violation}, who reported superconductivity in an interface-free, quasi-two-dimensional NiBi$_3$ thin film with thickness of approximately 6 nm on the top of MgO and Al$_2$O$_3$ substrates. In contrast to Bi/Ni bilayers, where interfacial effects and inversion-symmetry breaking play a central role, these samples consist of a continuous polycrystalline NiBi$_3$ layer, allowing the intrinsic superconducting properties of NiBi$_3$ to be probed in the two-dimensional limit. Magnetotransport measurements reveal a pronounced anisotropy in the superconducting response to magnetic fields, as shown in Fig.~\ref{fig:unconventional_nibi3}\textcolor{blue}{c,d}, where the superconducting transition is rapidly suppressed by out-of-plane fields but remains remarkably robust under in-plane magnetic fields up to several tesla.

Most remarkably, the in-plane upper critical field $B_{C2}^{\parallel}$ systematically exceeds the Clogston–Chandrasekhar spin-paramagnetic (Pauli) limit $B_P$ by a factor of 1.5, as summarized in Fig. \ref{fig:unconventional_nibi3}\textcolor{blue}{e}. This B$_P$ arises from Zeeman splitting of BCS spin-singlet Cooper pairs and defines the maximum field sustainable against paramagnetic pair breaking at T = 0 K \cite{chandrasekhar1962note,clogston1962upper}. Furthermore, conventional mechanisms capable of relaxing the Pauli limit were carefully examined. One possibility is spin–orbit scattering, where spin-flip processes induced by impurity scattering randomize the electron spin orientation and thereby reduce the effectiveness of Zeeman depairing enhancing B$_P$ \cite{klemm1975theory}. However, quantitative analysis shows that unrealistically short spin–orbit scattering times would be required to reproduce the experimental $B_{C2}^{\parallel}$, rendering this mechanism insufficient. Another scenario involves Rashba-type SOC, where spin–momentum locking admixes singlet and triplet components, partially protecting the superconducting state against Zeeman splitting. Nevertheless, this mechanism imposes a well-defined upper bound, modifying the paramagnetic limit to $B_P^*=\sqrt{2}B_P$ at T = 0 K \cite{gor2001superconducting,liu2018interface}. As highlighted in the dark-gray shaded region in Fig. \ref{fig:unconventional_nibi3}\textcolor{blue}{e}, the experimental data lies beyond this theoretical maximum, indicating that Rashba SOC alone cannot fully account for the robustness of superconductivity under in-plane magnetic fields.

Alternatively, detection of anomalous Hall–like signal close to the superconducting states in NiBi$_3$ thin film were consistently observed in refs. \cite{}. and particularly interpreted as ferromagnetic spin fluctuations persisting in low-dimensional NiBi$_3$. Similar to and robustness were reported in the work of Herrmestred for nanostructured NiBi$_3$ \cite{herman2012molecular}. Although a consisten explanation remains speculative, the reduction of it provides a physically motivated scenario in which dynamic magnetic correlations, enhanced by reduced dimensionality due to high surface-to-volume ratio, may promote an unconveitonal superocnductivity. Particularly,  singlet-to-triplet conversion or odd-frequency equal-spin triplet correlations. In this context, quasi-two-dimensional NiBi$_3$ emerges as a promising platform where intrinsic superconductivity, strong fluctuations, and reduced dimensionality conspire to produce superconducting behavior that departs from the conventional BCS paradigm.

\subsection{The origin of the superconductivity}

\begin{figure}
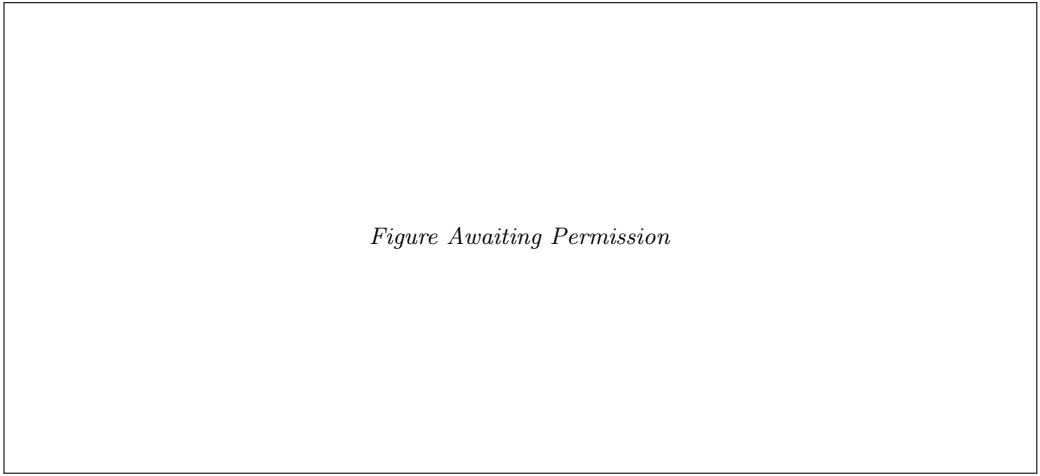

    \centering
    \fbox{
        \parbox[c][6cm][c]{0.95\linewidth}{
            \centering
            \textit{Figure Awaiting Permission}
        }
    }
    \caption{\textbf{Theoretical Model for Superconductivity in Bi/Ni Bilayers.} 
    \textbf{Upper panel}: Schematic representation of the system, highlighting the Bi-rich region with metallic surface states, the Bi$_3$Ni-rich region, and the Ni layer. The red arrow corresponds to the in-plane magnetization of Ni, while the black arrows indicate the induced exchange field. \textbf{Bottom panel}: Phase diagram of the proximity-induced superconducting gap $E_g/\Delta$ at Bi surface as a function of the chemical potential normalized by the Zeeman field ($\mu/h$) and the ratio between Rashba and warping spin-orbit coupling ($\lambda_R/\lambda_D$). The regions correspond to a trivial superconducting state (NSC), a topological $p \pm ip$ state (TSC), and a gapless regime. The yellow arrow marks the parameter set where the model predicts the maximal topological gap ($\sim 0.4$ meV). Data from \cite{chao2019superconductivity}.}
    \label{fig:review_bini_chaomodel}
\end{figure}

Essentially, the origin of superconductivity seems related to the deposition technique. For epitaxial growth, FCC-induced cubic Bi structures or amorphous Bi clusters have been argued to be responsible for superconductivity, while for non-epitaxial samples, the spontaneous formation of intermetallic NiBi$_3$ at the interface seems to dominate the superconducting transition. Still, a microscopic understanding remains elusive. In fact, one microscopic theoretical model has been developed, and surprisingly it attempts to include all these features \cite{chao2019superconductivity}. It consider the following three elements: (i) the ferromagnetic Ni layer, which generates an exchange field close to the interface; (ii) a NiBi$_3$ layer respossible by the superconductivity gap due to proximity effect; and (iii) the metallic surface state from a Bi layer \cite{xiao2012bi}, which introduce both Dresselhaus and Rashba SOC. This strong SOC locks spin and momentum, so that a conventional singlet pairing induced by proximity is inevitably mixed with triplet components. This model closely resembles the physics of long-range triplet generation at superconductor/ferromagnet interfaces with non-collinear magnetization, where conventional singlet pairs are converted into equal-spin triplets that can propagate large distances \cite{bergeret2001long,eschrig2007symmetries,bergeret2005odd}. Fig. \ref{fig:review_bini_chaomodel}\textcolor{blue}{b} shows the calculated phase diagram of the effective superconducting gap $E_g/\Delta$. Three regimes emerge from the phase diagram: (i) a trivial superconducting state (NSC), consistent with conventional NiBi$_3$ proximity effect; (ii) a gapless region, where quasiparticle bands close before reopening; and (iii) a topological $p \pm ip$ superconducting phase (TSC) with a finite topological gap. The yellow arrow indicates the parameter set where the model predicts the maximal topological gap ($\sim 0.4$ meV). Importantly, this point was not chosen arbitrarily: it corresponds to the set of parameters extracted from point-contact Andreev reflection spectroscopy, showing that the experimental data of ref. \cite{gong2015possible} place the Bi/Ni system inside the topological $p \pm ip$ phase.

The model provides a possible framework to reconcile the apparent dichotomy in Bi/Ni bilayers, as it captures both the unconventional superconducting features, such as the TRSB detection and the conductance spectra from Andreev spectroscopy, and the more conventional bulk-like behavior associated with NiBi$_3$. Nevertheless, a unified understanding has not yet been reached. In particular, for epitaxial samples no evidence of NiBi$_3$ formation has been reported, and in those cases the discussion is instead centered on the stabilization of FCC Bi structures or amorphous clusters as the source of superconductivity

\section{Gaps in the literature}

Although the fundamental aspects of NiBi$_3$ superconductivity points to a conventional $s$-wave BCS, several open questions remain regarding its normal-state transport and possible magnetic contributions. In bulk single crystals, state-of-the-art techniques have consistently reported the absence of long-range ferromagnetism, yet revealed surface-confined magnetic fluctuations whose microscopic origin is still debated. These observations contrast sharply with the pronounced ferromagnetic-like signatures in confined geometries, suggesting a dimensionality-driven instability that remains unexplained. From a transport perspective, NiBi$_3$ displays a number of unconventional features. The temperature dependence of $\rho_{xx}(T)$ deviates significantly from the standard Bloch–Grüneisen behavior, showing early resistivity saturation. In addition, magnetoresistance measurements have revealed violations of Kohler’s rule above a given temperature, implying an additional scattering channels beyond simple orbital effects. Despite these intriguing anomalies, systematic magnetotransport studies in high-quality single crystals are poor reported. 

As for the Bi/Ni bilayer case, the microscopic origin of superconductivity remains unsettled. In epitaxial films with Bi growth at cryogenic temperature, the transition has been linked to metastable phases of Bi, while in polycrystalline growth at room temperature the spontaneous formation of NiBi$_3$ appears to dominate. This contrast shows that the mechanism is highly sensitive to growth technique, as well, as the temperature conditions.  Moreover, the symmetry of the superconducting order parameter is also debated. Kerr effect, Andreev and Terahertz spectroscopy suggest unconventional chiral states, which strongly constant with the policrystalline case, where the order parameter follows the BCS predictions of $s$-wave pairing symmetry. Whether Bi/Ni hosts a genuine topological phase or simply reflects conventional superconductivity mediated by NiBi$_3$ is still an open question. Additionally, the role of spin-orbit coupling has yet to be firmly established, as it appears to be present in the case of non-epitaxial samples.

\section*{References}

\bibliography{aipsamp}

\end{document}